\documentclass{ws-ijmpcs}

\begin{document}

\markboth{J. G. Coelho and M. Malheiro}
{Similarities of SGRs with low magnetic field and white dwarf pulsars}

%
\catchline{}{}{}{}{}
%

\title{SIMILARITIES OF SGRs WITH LOW MAGNETIC FIELD AND WHITE DWARF PULSARS}

\author{J. G. COELHO AND M. MALHEIRO}

\address{Departamento de F\'{i}sica, Instituto Tecnol\'{o}gico de Aeron\'{a}utica, Pra\c{c}a Marechal Eduardo Gomes, 50 - Vila das Ac\'{a}cias\\
S\~{a}o Jos\'{e} dos Campos, 12.228-900,
Brazil\\
jaziel@ita.br\\
malheiro@ita.br}

%

\maketitle

\begin{history}
\received{Day Month Year}
\revised{Day Month Year}
\end{history}

\begin{abstract}
Some of the most interesting types of astrophysical objects that have been intensively studied
in the recent years are the Anomalous X-ray Pulsars (AXPs) and Soft Gamma-ray Repeaters (SGRs) seen
usually as neutron stars pulsars with super strong magnetic fields. However, in the last two
years two SGRs with low magnetic fields have been detected. Moreover, fast and
very magnetic white dwarf pulsars have also been observed in the last years. Based on these new pulsar
discoveries, white dwarf pulsars have been proposed as an alternative explanation to the
observational features of SGRs and AXPs. Here we present several properties of these SGRs/AXPs as
WD pulsar, in particular the surface magnetic field and the magnetic dipole momentum.

\end{abstract}

\ccode{PACS numbers: 11.25.Hf, 123.1K}

\section{Introduction}	
Recently, an alternative description of the Soft
Gamma-ray Repeaters (SGRs) and Anomalous X-ray Pulsars (AXPs)\cite{Mereghetti,Mereghetti2}
based on rotating highly magnetized and very massive white dwarfs (WD) has been proposed by
Malheiro, Rueda and Ruffini\cite{MMalheiro}. In this new description several observational
properties are easy understood and well explained as a consequence of the large radius of a massive
white dwarf that manifests a new scale of mass density, moment of inertia and rotational energy in
comparison with the case of neutron stars (NS).

In the last years, two SGRs with low magnetic field were discovered, SGR 0418+5729
and Swift J1822.3-1606 (see N. Rea et al.\cite{NandaRea,NandaRea2}). This sources were
found to have a period of $\sim9.08$ s and $\sim8.44$ s respectively.
These new astronomical observations challenged the
magnetar model of SGRs where the large magnetic field is
the source of the steady X-ray luminosity observed, and also responsible for the outbursts seen as
the main characteristic of these type of pulsar. Moreover, the characteristic age $\tau=P/2\dot{P}$
of these two magnetars are large with $\tau \sim (10^6-10^7)$ years, older comparing with all the others
SGRs and AXPs $\tau \sim (10^3-10^4)$ years, an age usually seen as an indication for
the association of SGRs/AXPs as young neutron star produced on supernova explosions.

\section{SGRs with low B as white dwarf pulsars}

The magnetic field at the magnetic pole $B_p$ is related to the dipole magnetic
moment by,
\begin{equation}\label{magnetic_moment}
\mid\overrightarrow{m}\mid= \frac{B_p R^3}{2},
\end{equation}
where $R$ is the star radius. If the star magnetic dipole moment is misaligned with the spin axis
by an angle $\alpha$, electromagnetic energy is emitted at
a rate (see, e.g., Shapiro and Teukolsky\cite{shapiro} and references therein),
\begin{equation}\label{edot}
\dot{E}_{\rm dip}= -\frac{2}{3c^3}\mid \ddot{m}\mid^2= -\frac{2\mid\overrightarrow{m}\mid^2}{3c^3}\omega^4\rm sin^2\alpha.
\end{equation}

Thus, it is the magnetic dipole moment of the star, the physical quantity that
dictates the scale of the electromagnetic radiated power emitted together
with the angular rotational frequency. The fundamental physical idea of the rotation-powered
pulsar is that the X-ray luminosity - produced by the dipole field - can be expressed
as originated from the loss of rotational energy of the pulsar,
\begin{equation}\label{erot}
 \dot{E}_{\rm rot}=-4\pi^2I \frac{\dot{P}}{P^3},
\end{equation}
associated to its spin-down rate $\dot{P}$, where $P$ is the rotational period and $I$ is
the momentum of inertia.

Thus, equaling Eqs.~(\ref{edot}) and (\ref{erot}) we deduce the expression of pulsar
magnetic dipole moment,
\begin{equation}\label{moment}
 m=\left(\frac{3c^3I}{8\pi^2}P\dot{P}\right)^{1/2}.
\end{equation}

From Eq.~(\ref{magnetic_moment}) we obtain the magnetic field at the equator $B_e$ as
\cite{Ferrari&Ruffini_1969}
\begin{equation}\label{MagneticField}
 B_e=B_p/2=\left(\frac{3c^3I}{8\pi^2R^6}P\dot{P}\right)^{1/2},
\end{equation}
where $P$ and $\dot{P}$ are observed properties and the moment
of inertia $I$ and the radius $R$ of the object model dependent properties. 
The model commonly addressed as
magnetar\cite{Duncan,Thompson} is based on a canonical neutron star of $M=1.4M_{\bigodot}$ and $R=10$
km and then $I\sim 10^{45}\rm{g}$ $\rm{cm^2}$ as the source of SGRs and AXPs.
From Eqs.~(\ref{moment}) and (\ref{MagneticField}), using the parameters above, we obtain the
magnetic dipole moment and the magnetic field
of the neutron star,
\begin{equation}\label{m_NS}
 m_{\rm NS}=3.2\times10^{37}(P\dot{P})^{1/2} \rm{emu},
\end{equation}
and
\begin{equation}\label{B_NS}
 B_{\rm NS}=3.2\times10^{19}(P\dot{P})^{1/2} \rm{G}.
\end{equation}

For the case of the white dwarf model we use a
radius $R= 3000$ km for all SGRs and AXPs and a mass $M=1.4M_{\bigodot}$, as recent studies
of fast and
very massive white dwarfs obtained (see K. Boshkayev et al.\cite{JRueda}). Thus, these values
of mass and radius generating the momentum of inertia $I\sim 1.26\times 10^{50}\rm{g}$ $\rm{cm^2}$,
will be adopt hereafter in this work as the fiducial white dwarf model parameters.
Using that parameters we obtain the
magnetic dipole moment and the magnetic field
of the white dwarf pulsar,
\begin{equation}\label{m_WD}
 m_{\rm WD}=1.14\times10^{40}(P\dot{P})^{1/2} \rm{emu},
\end{equation}
and
\begin{equation}\label{B_WD}
 B_{\rm WD}=4.21\times10^{14}(P\dot{P})^{1/2} \rm{G}.
\end{equation}

The recent discoveries of
SGR 0418+5729 and Swift J1822.3-1606 with low magnetic field sharing some properties with the
recent detected fast WD pulsar AE Aquarii and RXJ 0648.0-4418, and the candidate EUVE J0317-855 is
also discussed in this paper (see Table~\ref{ta1}), to support the WD description of SGRs and AXPs
white dwarf pulsar.

Massive fast rotating white dwarfs with magnetic fields larger than $10^6 \rm{G}$
up to $10^9 \rm{G}$ has been observed\cite{Terada2,Angel}. A specific example is AE Aquarii,
the first white dwarf pulsar, very fast with a short period $P=33.08$ s\cite{Terada3,Terada4,Terada5}.
The rapid braking of the white dwarf and the nature of pulse hard X-ray emission
detected with SUZAKU space telescope in 2005 under these conditions can be explained in
terms of spin-powered pulsar mechanism\cite{Ikhasanov2,Ikhasanov3}.
Although AE Aquarii is a binary
system with orbital period $\sim9.88$ hr, very likely the power due to accretion of matter is
inhibited by the fast rotation of the white dwarf.

More recently, the X-ray multimirror mission (XMM)-Newton satellite had observed a white dwarf pulsar
faster than AE Aquarii. Mereghetti et al.\cite{Mereghetti3}
showed that the X-ray pulsator RX J0648.0-4418 is a white dwarf with mass $M=1.28M_\odot$ and radius
$R= 3000$ km. Is one of the most massive white dwarfs currently known and the one with
the shortest spin period $P = 13.2$ s\cite{Mereghetti3,Mereghetti4}. That belongs to the binary
systems HD 49798/RX J0648.0-4418.

As discussed by Mereghetti et al.\cite{Mereghetti3}, the luminosity of $L_X \sim 10^{32}$ erg/s is produced by accretion
onto the white dwarf of the helium-rich matter from the wind of the companion. In this work, we
do not consider the accretion model, instead we
describe RX J0648.0-4418 as
a rotation powered white dwarf, and obtain the magnetic dipole moment and magnetic field,
using the above mentioned parameters of a fast rotating magnetized
white dwarf.

EUVE J0317-855, a hydrogen-rich magnetized white dwarf discovered as an extreme-ultraviolet (EUV)
source by the {\it{ROSAT Wide Field Camera}} and {\it{Extreme Ultraviolet Explorer EUVE}} survey,
is another observed WD pulsar candidates\cite{Kulebi,Ferrario}. However, relevant pulse emission from
EUVE J0317-855 has not been observed yet, which may suggest that the electron-positron creation
and acceleration does not occur (see Kashiyama et al.\cite{Kashiyama}). Barstow et al.\cite{Barstow} obtained a period
of $P\sim$ 725 s, which is also a fast and very magnetic white dwarf with a dipole magnetic field is
$B\sim 4.5\times 10^8$ G, and a mass $(1.31 - 1.37)M_\odot$ which is relatively large
compared with the typical WD mass $\sim 0.6 M_\odot$. EUVE J0317-855 has a white dwarf companion, but is supposed to be
no interaction between them, because of their large separation ($\gtrsim 10^3$ UA)\cite{Kashiyama}.

\begin{table}[ph]
\tbl{Comparison of the observational properties of five sources: SGR 0418+5729 and Swift J1822.3-1606 (see N. Rea et al. 2010, 2012)
and three observed white dwarf pulsar candidates. For the SGR 0418+5729 and Swift J1822.3-1606 the parameters
$P$, $\dot{P}$ and $L_X$ have been taken from the McGill
online catalog at www.physics.mcgill.ca/~pulsar/magnetar/main.html. The characteristic age is given
by Age $=P/2\dot{P}$ and the magnetic moment $m$ and
the surface magnetic field $B$ are given by Eqs.~(\ref{moment}) and (\ref{MagneticField}) respectively.}
{\begin{tabular}{@{}cccccccc@{}} \toprule
 & SGR 0418+5729 &Swift J1822.3-1606 &AE Aquarii &RXJ 0648.0-4418 &EUVE J0317-855 \\
\\ \colrule
$P$(s) &9.08 &8.44 & 33.08 &13.2 &725\\
$\dot{P}$ $(10^{-14})$ &$<$ 0.6 &8.3 &5.64 &$<$ 90 &- \\
$\rm Age$ $(\rm Myr)$ &24 &1.6 &9.3 &0.23 &- \\
$L_X$ $(\rm erg/s)$ &$\sim6.2\times10^{31}$ &$\sim4.2\times10^{32}$ &$\sim10^{31}$ &$\sim10^{32}$ &- \\
$B_{\rm WD}$(G) &$<$ $9.83\times10^7$ &$3.52\times10^{8}$ &$\sim 5\times10^7$ &$<$ $1.45\times10^9$ &$\sim 4.5\times10^8$ \\
$B_{\rm NS}$(G) &$<$ $7.47\times10^{12}$ & $2.70\times10^{13}$ &- &- &- \\
$m_{\rm WD}$(emu) &$2.65\times10^{33}$ &$0.95\times10^{34}$ &$\sim1.35\times10^{33}$ &$3.48\times10^{34}$ &$1.22\times10^{34}$ \\
$m_{\rm NS}$(emu) &$7.47\times10^{30}$ &$2.70\times10^{31}$ &- &- &- \\\botrule
\end{tabular} \label{ta1}}
\end{table}

In Table~\ref{ta1} we compare and contrast the parameters of these two SGRs with low B described 
in the white dwarf model with the three fast white dwarfs presented before. The magnetic 
dipole moment and magnetic field of all the sources are calculated using
WD and NS fiducial parameters discussed above.

We see from Table~\ref{ta1} that several features of two SGRs with low magnetic field are
very similar to the ones of fast and magnetic white dwarfs recently detected. They are old,
characteristic ages of Myr, low quiescent X-ray luminosity $L_X \sim (10^{31}-10^{32})$,
magnetic field of $B_{\rm WD}\sim (10^7-10^8)$ G and magnetic dipole moments of $m_{\rm WD}\sim (10^{33}-10^{34})$ emu.
These results give evidence for the interpretation of SGRs/AXPs as being rotating white dwarf pulsars.

\section{Conclusions}

The values for $m\sim (10^{33}-10^{34})$ emu of the two SGRs with low B, are exactly at the same order
of the three white dwarf pulsars observed (see Table~\ref{ta1}), and in the lower values of the
observed isolated and polar white dwarf magnetic dipole moment range\cite{Terada5}.

The large steady X-ray emission $L_X\sim 10^{35}$ erg/s observed in the SGRs/AXPs now
well understood as a consequence of the fast white dwarf rotation ($P\sim 10$ s), since
the magnetic dipole moment $m$ is at the same scale as the one observed for the very magnetic 
and not so fast white dwarfs.
This supports the description of SGRs and AXPs as belonging to a class of very fast and magnetic
massive white dwarfs\cite{MMalheiro} perfect in line with recent astronomical observations
of fast white dwarf pulsars.

\section*{Acknowledgments}

The authors acknowledges the financial support of the Brazilian agency CAPES, CNPq
and FAPESP (S\~{a}o Paulo state agency, thematic project 2007/03633-3).

\end{document}